**TITLE:**

PROPERTIES OF A MONOLITHIC SAPPHIRE PARAMETRIC TRANSDUCER: PROSPECTS OF MEASURING THE STANDARD QUANTUM LIMIT

**AUTHORS:**

C.R. Locke[#], M.E. Tobar, E.N. Ivanov

University of Western Australia, Crawley, Australia. [#]Email: clocke@pd.uwa.edu.au

**SHORT TITLE:**

Prospects of measuring the standard quantum limit

**PACS:**

ABSTRACT:

To measure the Standard Quantum Limit (SQL) a high quality transducer must be coupled to a high quality mechanical system. Due to its monolithic nature, the Monolithic Sapphire Transducer (MST) has high quality factors for both types of resonances. Single loop suspension is shown to yield a mechanical quality factor of $6x10^8$ at 4K and Whispering Gallery electromagnetic modes display a quality factor of $2x10^6$ at 4K. From standard analysis we show that the MST has the potential to measure the noise fluctuations of the mechanical oscillator at the SQL. Also, we point out a new way to determine if the transducer back-action is quantum limited. We show that if the fluctuations are at the quantum limit, then the amplitude of the oscillations will be amplified by the ratio of the ringdown time to the measurement time, which is an inherently easier measurement.

## 1. <u>Introduction</u>

### *1.1. Monolithic sapphire transducer*

The transducer developed in this work is a cylinder of diameter 50mm and length 100mm, effective mass of 0.4kg (half the actual mass) and resonant frequency of 53kHz. It is manufactured from high purity HEMEX grade [1] sapphire, having low electrical and mechanical losses due to its single-crystal nature. We have called this the Monolithic Sapphire Transducer (MST.)

An internal electromagnetic resonance is excited within the MST, as shown in Figure 1. How well the MST will retain the internal electromagnetic resonance is quantified by the electrical "quality factor" (Q-factor.) Similarly, the mechanical Q-factor quantifies the acoustic energy retention. The Q-factor is defined as the ratio of the stored energy associated with the resonance to the energy dissipated per cycle.

Sapphire has a low dielectric constant (of the order of 10), resulting in large radiation losses and difficulty in trapping energy within a sapphire oscillator. This is overcome through using a class of resonant microwave frequency modes called "Whispering Gallery" (WG) modes. These modes comprise of standing electromagnetic waves within the surface of the resonator having a high confinement due to the shallow angle of reflection from the surface.

Frequency modulation of WG modes occurs when the sapphire bar oscillates in its fundamental acoustic mode of vibration, causing dimensional changes to the crystal (as shown in Figure 1.) This leads to a twofold mechanism of modulation of the WG resonances; firstly, the dimension change itself changes the boundary conditions of the WG resonance, and secondly, the permittivity changes due to strain induced by the dimension change. The latter is the dominant effect. This can be thought of as a resonant inductor/capacitor/resistor circuit coupled to a mass/spring system, in which the oscillating mass changes the capacitance of the circuit [2].

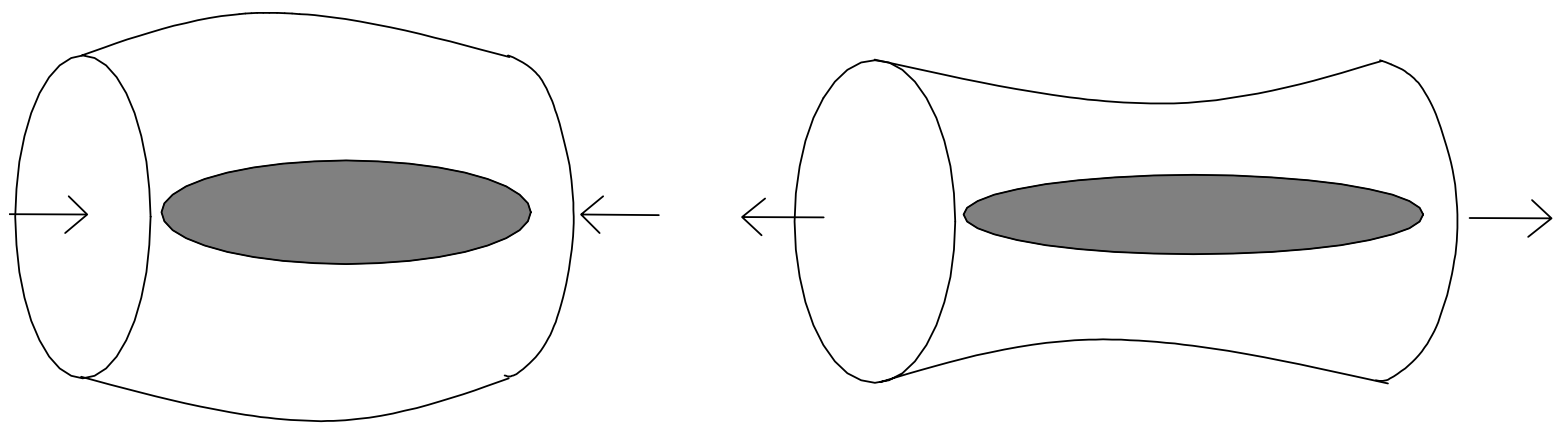

*Figure 1: The MST cylinder oscillating in its acoustic mode of vibration will change the frequency of the internal electromagnetic resonance (shown as the dark region.)*

Using the well-known Nyquist formula of the spectral density of force the expected oscillator displacement and noise variance is calculated. We then study the factors limiting the current sensitivity of the transducer (predominately the phase noise of the pump source) and methods to reduce these noise sources. Results of the improved suspension system, which utilizes a single loop of wire, are presented. Further calculations then allow us to determine the requirements to measure quantum mechanical effects in a macroscopic mass.

### *1.2. Standard Quantum Limit for an acoustic oscillator*

As quantified in Heisenberg's uncertainty relation, there are fundamental restrictions on the accuracy of macroscopic measurements. We follow a slightly alternative approach than [3] in the theoretical analysis of quantum noise. Consider a harmonic oscillator of mass m (Figure 2) driven by a random force $F_N(t)$. The response is intrinsically narrow band and can be represented as a sinusoidal displacement $x$ and a random component $\Delta x$.

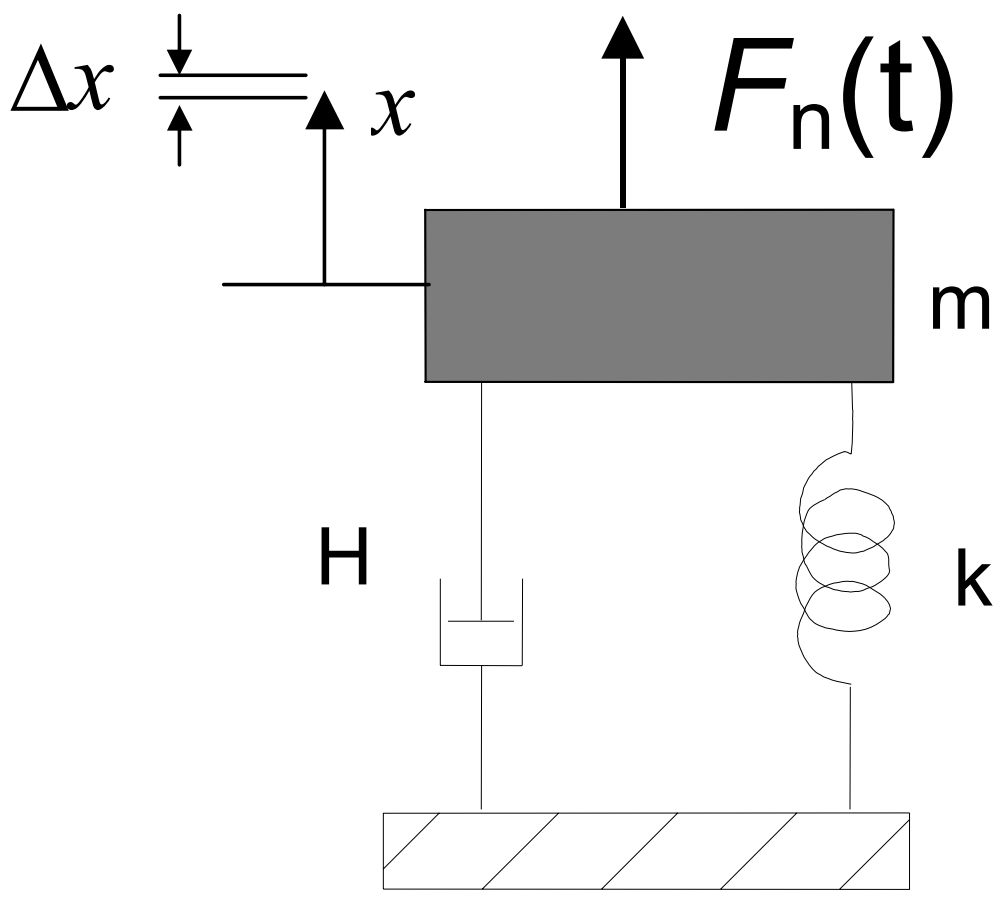

*Figure 2: Harmonic oscillator driven by a random force. H is the damping constant and k the spring constant.*

The equation of motion can be written as

$$m\frac{d^2x}{dt^2} + H\frac{dx}{dt} + kx = F_N(t)$$

*Equation 1*

where k and H are the spring and damping constants of the mechanical oscillator, respectively. Taking the Fourier transform, and using

$$\omega_0^2 = \frac{k}{m} \qquad and \qquad Q = \frac{\omega_0 m}{H}$$

where $\omega_0$ is angular resonant frequency and Q the mechanical quality factor, yields;

$$X^2(\omega) = m^{-2}\omega_0^{-4}G^2(\omega)S_F(\omega) \quad \text{where} \quad G^2(\omega) = \left(\left(1 - \frac{\omega^2}{\omega_0^2}\right) + \left(\frac{\omega}{Q\omega_0}\right)^2\right)$$

*Equation 2*

Here $G(\omega)$ is the magnitude of the displacement gain of the resonance, $S_F(\omega)$ the spectral density of the force excitation [$N^2$/Hz], and $X^2(\omega)$ the spectral density of displacement fluctuations [$m^2$/Hz].

Assuming $S_F$ is constant means

$$\left\langle x^2 \right\rangle = 2\int_0^\infty X^2(\omega)d\omega = \frac{2S_F}{m^2\omega_0^4}\int_0^\infty G^2(\omega)d\omega$$
$$= \frac{S_F\tau}{4m^2\omega_0^2}$$

*Equation 3*

Over large averaging times the oscillator will be in thermal equilibrium ($t_{meas} > \tau$), where $t_{meas}$ is the measurement time, and $\tau$ is the characteristic decay time of the oscillator. In this case the mean displacement $<x>^2$ of a sinusoidal oscillation is zero, so the variance of the oscillator may be calculated from,

$$\langle x^2 \rangle = \langle x \rangle^2 + (\Delta x)^2 \qquad \Rightarrow \qquad (\Delta x)^2 = \frac{S_F \tau}{4m^2 \omega_0^2}$$

*Equation 4*

However, if the measurement time is smaller than $\tau$, the oscillator's random motion is decoupled from the thermal bath (ie. measurement is performed quicker than the average fluctuations) and the displacement variance is filtered by the factor $t_{meas}/\tau$, and becomes,

$$(\Delta x)^2 = \langle x^2 \rangle \frac{t_{meas}}{\tau} = \frac{S_F t_{meas}}{4m^2 \omega_0^2}$$

*Equation 5*

### *1.2.1 Nyquist Noise*

Assuming that $S_F$ is driven by the thermal Nyquist force, given by

$$S_{F_{NYQ}} = \hbar \omega_0 H \coth\left(\frac{\hbar \omega_0}{2k_b T}\right)$$

*Equation 6*

where T is temperature and $k_b$ Boltzmann's constant. Then when $T > \hbar \omega_0 / 2k_b$,

$$(\Delta x)^2_{NYQ} = \langle x^2 \rangle_{NYQ} = \frac{k_b T}{m \omega_0^2} \qquad when \ \ t_{meas} > \tau$$

$$(\Delta x)^2_{NYQ} = \langle x^2 \rangle_{NYQ} \frac{t_{meas}}{\tau} = \frac{k_b T}{m \omega_0^2} \frac{t_{meas}}{\tau} \qquad when \ \ t_{meas} < \tau$$

*Equation 7*

For $t_{meas}<\tau$ we can define the effective temperature;

$$T_{eff_{NYQ}} = T\left( {}^{t_{meas}}\!/_{\tau} \right)$$

*Equation 8*

At small measurement times, Equation 7 can be expressed as;

$$(\Delta x)^2_{NYQ} = \frac{S_{Feff_{NYQ}} \tau}{4m^2 \omega_0^2}$$

*Equation 9*

where

$$S_{Feff_{NYQ}} = {}^{4mk_B T_{eff_{NYQ}}}\!/_{\tau}$$

In the case when the temperature is in the regime $T < \hbar\omega_o/(2k_B)$ then Equation 6 reduces to the Nyquist quantum spectral density of force [$N^2$/Hz];

$$S_{F_{NYQ}} = \hbar\omega_0 H = 2\hbar\omega_0 m / \tau$$

*Equation 10*

Substituting Equation 10 into Equation 4 we determine that

$$(\Delta x)^2_{SQL_{NYQ}} = \langle x^2 \rangle_{SQL_{NYQ}} = \frac{\hbar}{2m\omega_o} \quad when \quad t_{meas} > \tau$$

*Equation 11*

This equation, derived from the quantum spectral density of force when $t_{meas} > \tau$, yields the well-known standard quantum limit for displacement. However, this Nyquist component will be filtered in a similar way to the thermal noise given by Equation 5 when $t_{meas} < \tau$. This implies that the narrow band noise due to the quantum fluctuations may be reduced arbitrarily by taking smaller measurement times. However, this is not true if one considers the back action of quantum fluctuations introduced by the measurement system.

### *1.2.2 Quantum back action noise*

To calculate the spectral density of force acting back on the system due to the measurement process, the force density ($F^2 = 2\langle\Delta p\rangle^2$ in [N/Hz]) created by a momentum perturbation must be considered. To calculate the spectral density of force this must be multiplied by the bandwidth over which the perturbation occurs. This is twice the minimum measurement bandwidth, as the perturbation will act over both positive and negative frequencies offset from the probe oscillator carrier frequency.

$$S_{F_{PERT}} = 4\langle\Delta p\rangle^2 / t_{meas}$$

*Equation 12*

The optimal momentum perturbation due to a quantum limited measurement can be calculated to be $\langle\Delta p\rangle^2 = \hbar\omega_o m/2$. Substituting this value into Equation 12, the optimal spectral density of force acting on the probe oscillator can be calculated to be;

$$S_{F_{PERT}} = 2\hbar\omega_o m / t_{meas}$$

*Equation 13*

By equating the quantum spectral density of force given by Equation 13 to the thermal spectral density of force given by Equation 6, one can calculate that the oscillator will behave quantum mechanically when

$$\frac{2k_B T_{eff}}{\omega_0} \leq \hbar \quad T_{eff} = T\left( {t_{meas}}/{\tau} \right)$$

*Equation 14*

Now by substituting Equation 13 into Equation 5 one can determine that;

$$(\Delta x)^2_{SQL_{PERT}} = \frac{\hbar}{2m\omega_o} \quad when \quad t_{meas} < \tau$$

*Equation 15*

The SQL of displacement is independent of the measurement time. However, if $t_{meas} < \tau$;

$$\left\langle x^2 \right\rangle_{SQL_{Pert}} = \frac{\hbar}{2m\omega_o} \frac{\tau}{t_{meas}}$$

*Equation 16*

The evolution of the thermal and quantum amplitudes will be equivalent as the same underlying physics explains both systems. However, as the spectral density of force due to quantum effects increases with decreasing measurement time, the mean square displacement of the quantum oscillator will increase (Equation 16), while the displacement variance remains constant (Equation 9.) This is a very significant point. It means the amplitude of the oscillations due to quantum back-action may be measured above the SQL by a factor $\tau/t_{meas}$, which can be much larger than noise fluctuations.

The converse applies for thermal amplitudes, as the spectral density of force is independent of measurement time. The mean square displacement remains constant while the displacement variance decreases proportionally to the measurement time. The effective spectral density of force fluctuations $S_{Feff}$ that limits the measurement when $t_{meas} < \tau$ can be determined by equating Equation 9 and Equation 15, to obtain;

$$S_{Feff_{PERT}} = S_{F_{PERT}} \frac{t_{meas}}{\tau} = {2\hbar\omega_o m}/{\tau}$$

*Equation 17*

which is equivalent to Equation 10.

## 2. Experiment

### *2.1. Experimental setup*

When the cylindrical MST is oscillating in its fundamental mode of vibration (53kHz) there is a minimum displacement at the centre of the cylinder, therefore suspension at this point is an obvious choice. The MST is suspended by a single loop of wire around the midpoint of the cylinder. The wire is in turn attached to a second stage of vibration isolation consisting of a two stage mass-spring system of copper masses (mass 2kg each) and low frequency springs.

The MST and suspension arrangement were lowered into a dewar and then cooled to cryogenic temperatures. Excitation of the acoustic resonance was by means of a mechanical relay switch that strikes the sapphire upon an end face. Microwave resonances are excited by a probe positioned to give an electrical coupling of approximately 0.2 at room temperature.

The microwave pump source consists of a liquid-nitrogen cooled sapphire oscillator (of frequency 8.96GHz) mixed with the signal from a HP8662A Synthesized Signal Generator to provide a stable, tunable source. After suitable filtering the input signal was amplified using a high power MITEQ amplifier capable of an output power of up to 1W for frequencies in the microwave X-band. The mechanical resonance adds sidebands to the reflected signal at plus and minus the acoustic angular frequency ($\pm\omega$) offset from the microwave carrier. The signal is then mixed with a phase-shifted portion of the carrier signal, which is adjusted such that the system is phase sensitive. The sapphire is then mechanically excited and from observation of the time evolution of the sidebands added to the carrier frequency the mechanical quality factor can be determined. This is done using a computer connected to the readout system via a GPIB interface.

### *2.2. Noise components*

To achieve a measurement of the SQL, it must be measurable over the other sources of noise in the system. These include Nyquist (thermal) noise, seismic noise, and electronic noise in the readout system. Seismic noise can be reduced using a suitable vibration isolation system. The electronic noise comprises of phase noise from the pump source and noise from microwave components. The pump source is created by mixing a liquid nitrogen cooled sapphire oscillator with a HP8662A frequency generator. Although the sapphire oscillator has a very low Single Side-Band (SSB) phase noise of order -165dBc/Hz the overall performance of the combined system is dominated by the comparably poor performance of the HP8662A frequency generator, having phase noise of –137 dBc/Hz. The displacement sensitivity limit due to this phase noise is of the order of 4 x $10^{-15}$ m/√Hz.

A further contribution to broadband noise in the system comes from the mixer in the readout. The level of mixer noise during measurement is approximately 5 x $10^{-15}$m/√Hz. This poor performance does not affect the measurements presented in this paper; however, to reach the SQL this must be further reduced. This can be done in a straightforward manner by adjusting the power entering the RF port of the mixer, and, if state-of-the-art interferometric readout systems are employed [4], we have the potential to reduce the level of readout noise to less than $10^{-19}$ m/√Hz.

Thermal noise is a more fundamental limit, as expressed in Equation 7. To determine the height of the peak in spectral domain ($S_x(\omega_o)$), the mechanical Q-factor must be taken into account; the higher the mechanical Q-factor the narrower and higher the thermal noise curve.

$$S_x(\omega_o) = \sqrt{\frac{4k_b T_0 Q_m}{m\omega_o^3}}$$

*Equation 18*

where $Q_m$ is the mechanical Q-factor. For our system at room temperature and with $Q_m$ = $1.2\text{x}10^8$ we find $S_x(\omega=\omega_o) = 1.2\text{x}10^{-14}$ m/√Hz.

Measured displacement at room temperature for the MST is shown in Figure 3; the thermal noise peak is just above the noise floor at $10^{-14}$ m/√Hz, which is of the same order as expected. The measured broadband noise floor of $10^{-14}$ m/√Hz corresponds to the calculated contributions of phase noise of the pump and mixer noise.

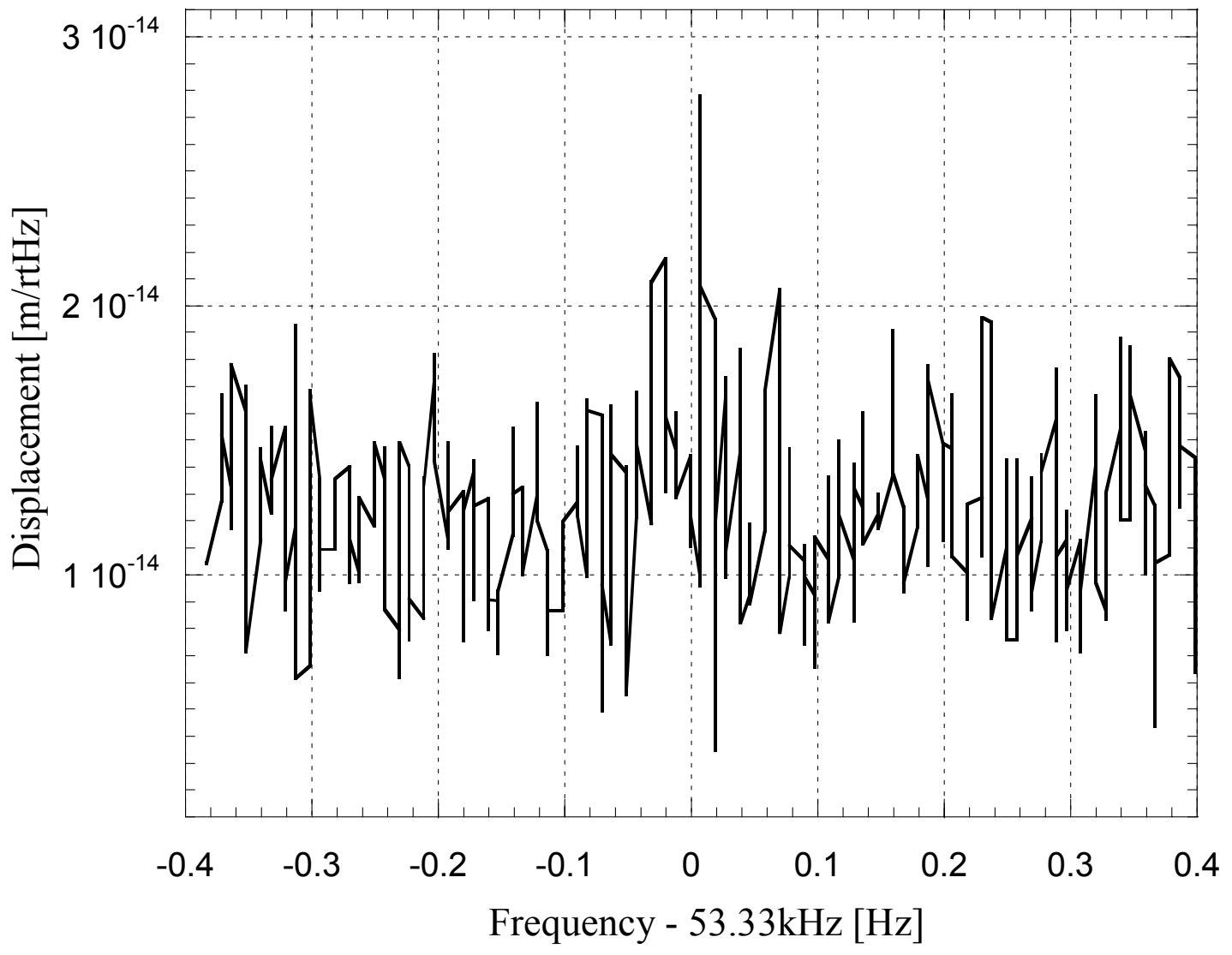

*Figure 3: Noise floor at room temperature, with a noise peak at the resonant frequency slightly above the broadband noise set by the mixer and phase noise of the pump.*

To reach the SQL the thermal noise in the MST must be reduced to the point where the following ratio is less than one;

$$\frac{\Delta x_{th}}{\Delta x_{SQL}} = \sqrt{\frac{k_b T_o t_{meas}}{\hbar Q_m}}$$

*Equation 19*

Hence, the temperature $T_o$ must be reduced, the mechanical Q maximized, and the measurement bandwidth maximised (ie. minimise measurement time.)

This experiment is designed to operate in a liquid helium environment, so a minimum temperature is about 2 K. Two processes limit the measurement bandwidth; (i) the Nyquist sampling theory, which requires $1/t_{meas} \leq \omega_o$, and (ii) the level of electronic series noise added by the readout.

The detectable displacement noise due to the readout is given by

$$\Delta x_{ro} = \sqrt{\frac{S_x(\omega_o)}{t_{meas}}}$$

*Equation 20*

where $S_x(\omega_o)$ [m$^2$/Hz] is the spectral noise density of the readout electronics. In this case

$$\frac{\Delta x_{ro}}{\Delta x_{SQL}} = \sqrt{\frac{2mS_x(\omega_o)\omega_o}{\hbar\, t_{meas}}}$$

*Equation 21*

Thus to measure the SQL above the series noise, one requires a small mass and frequency and large measurement time. In practice there is an optimum measurement time that balances Equation 19 and Equation 21.

An easier way to determine if the narrow band noise is quantum limited is to measure the amplitude of the mechanical oscillator as a function of the measurement time $t_{meas}$. If quantum limited, the amplitude will be dependent on $t_{meas}$, as shown in Figure 4.

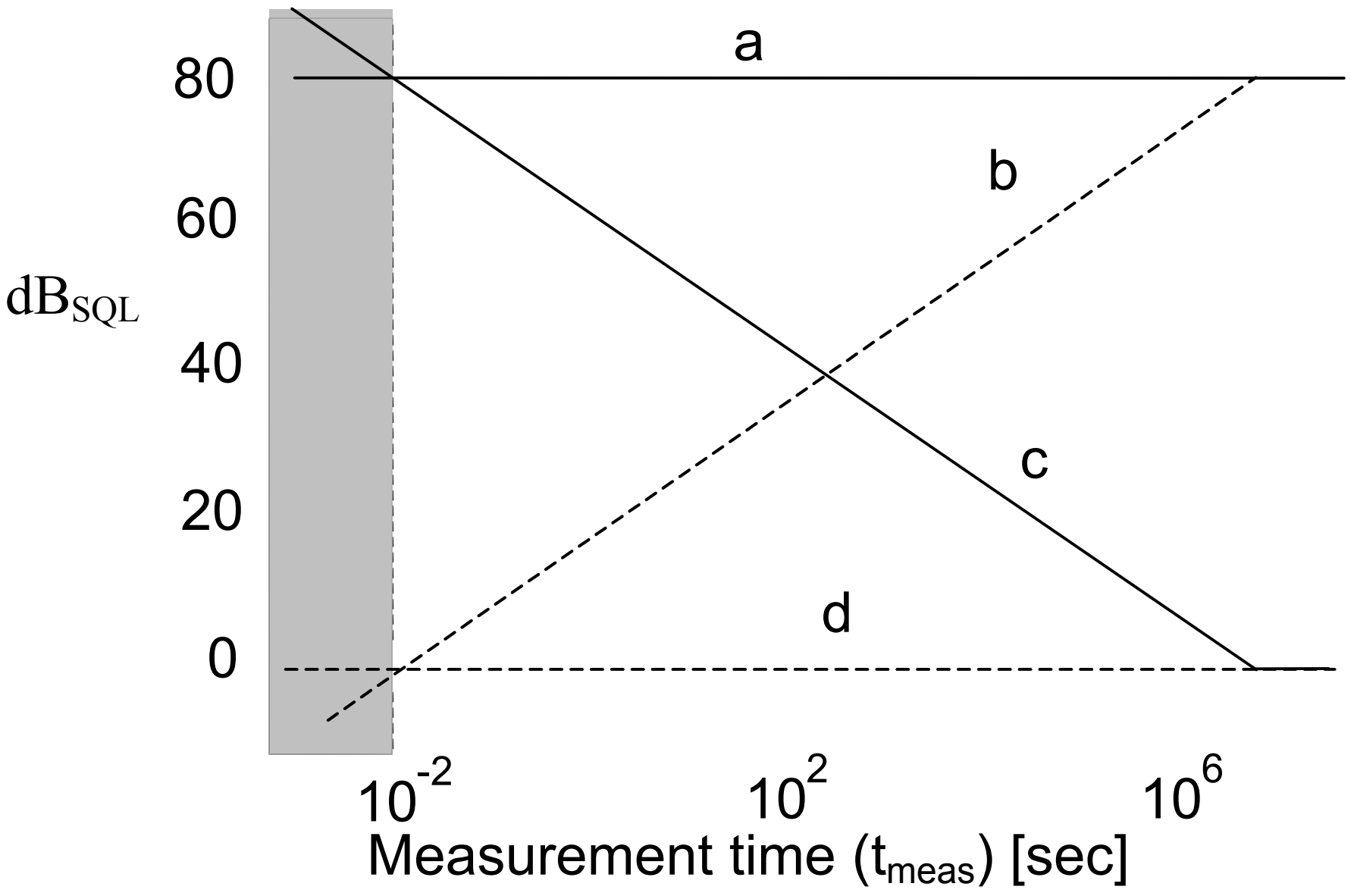

*Figure 4: The mean square displacement (bold) and variance (dashed) for the thermal (a & b) and quantum noise (c & d) as a function of measurement time for a sapphire probe oscillator. The shaded region shows the measurement times when the quantum back-action amplitude is enhanced (Equation 16.) In gravitational wave detection it is common to express amplitude in units of temperature. In this case a and c are known as mode temperatures, and b and d are known as effective noise temperatures.*

Here, if the measurement time is sufficiently reduced, the quantum back-action component of the amplitude will be above the Nyquist amplitude.

### *2.3. Mechanical Q-factor*

Experiments were conducted that compared the effectiveness of this single loop of wire against niobium "Catherine Wheel" type suspension systems. The Catherine Wheel [5] is a cantilever spring suspension, essentially a scaled down version of the niobium flexure supporting the gravitational wave antenna "Niobe" at UWA. The maximum mechanical Q-factor of the MST at room temperature attained using the Catherine Wheel was $5 \times 10^6$. Even this comparatively poor performance was not reliably reproducible due to inevitable plastic deformation of the Catherine Wheel each time it is loaded.

Single loop wire suspension, however, proved more stable, cheaper to construct, and produced superior results.

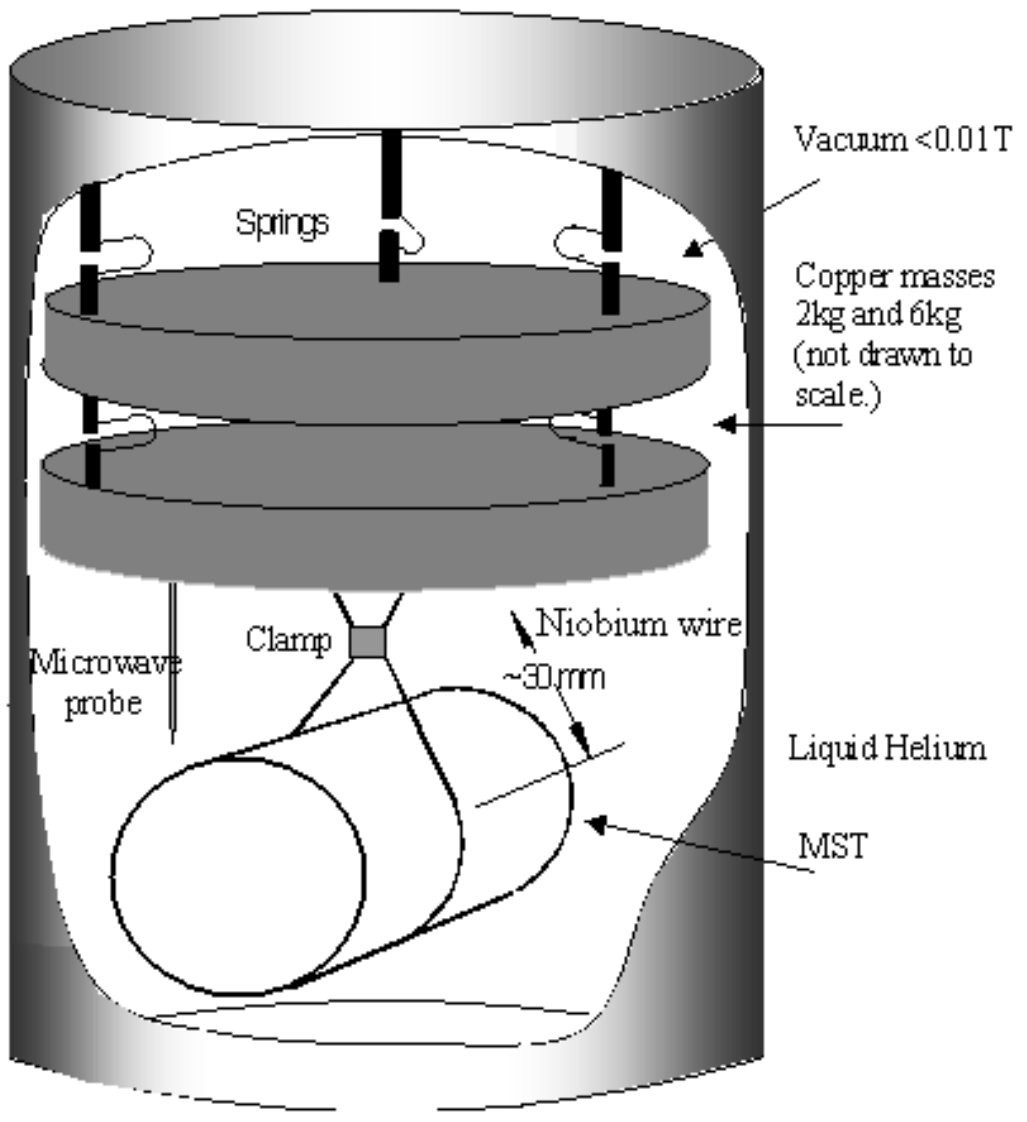

*Figure 5: Experimental arrangement for single loop suspension of the MST in a liquid helium environment.*

Various materials were tested for use as the wire suspension, including nylon, tantalum, tungsten, and niobium.

| *Material* | *Nylon* | *Tantalum* | *Tungsten* | *Niobium* | *Niobium* |
|---|---|---|---|---|---|
| Wire diameter | 800μm | 200μm | 110μm | 150μm | 125μm |
| Best mechanical Q-factor | $1.5x10^7$ | $4x10^6$ | $6x10^7$ | $5x10^7$ | $10^8$ |

*Table 1: Single loop suspension*

Significant losses occur when the internal mode of the suspension system couples to the suspended sample. When the resonating sample is well coupled to the wire suspension, the wire will be excited and elastic waves will dissipate energy from the system. An equation for the losses in a suspended resonator can be derived [6],

$$\frac{1}{Q_m} = \frac{2l\rho A x_{ext}^2}{MQ_w x_0^2}\left(1 + \frac{1}{2}\left(\frac{\omega l}{\upsilon Q_w}\right)^2 - \cos\left(\frac{2l\omega}{\upsilon}\right)\right)^{-1}$$

*Equation 22*

Here $Q_m$, $Q_w$ are the mechanical quality factors of the resonator and wire, *l* the length of the wire, ρ its density, A the cross sectional area, and ν the speed of the travelling waves in the wire. M is the effective mass and ω the resonant frequency of the resonator. $x_0$ is the amplitude of vibration of the resonator and $x_{ext}$ the amplitude at point of contact between the resonator and the wire. The ratio $x_0/x_{ext}$ is estimated from finite element modelling to be 0.2. We found agreement between theory and results, the experimental measurements displaying predictable periodic behavior, yielding an optimum length of wire of 37mm.

The surface of sapphire in contact with the wire was found to be important, as noted by Braginsky [6]. We obtained best results when a thin layer of grease covered the section of sapphire in contact with the wire.

| *Wire length (mm)* | *35.5* | *36.5* | *37.5* | *38.0* |
|---|---|---|---|---|
| Clean | $1.0\ 10^7$ | $2.0\ 10^7$ | $1.0\ 10^8$ | $2.9\ 10^7$ |
| Grease layer | $5.5\ 10^7$ | $9.0\ 10^7$ | $1.5\ 10^8$ | $9.5\ 10^7$ |
| Factor improvement | 5.5 | 4.5 | 1.5 | 3.2 |

*Table 2: Effect of a layer of grease on mechanical Q-factor*

The excellent performance of niobium wire has been demonstrated over the range from room to liquid helium temperatures, as shown in Figure 6. At room temperature a maximum mechanical Q-factor of $1.4 \times 10^8$ for the MST was measured. After the MST was cooled to 4K the mechanical Q-factor improves to $6 \times 10^8$. From analysis of the noise mechanisms we believe that at 4K the intrinsic Q-factor of this piece of sapphire has been reached.

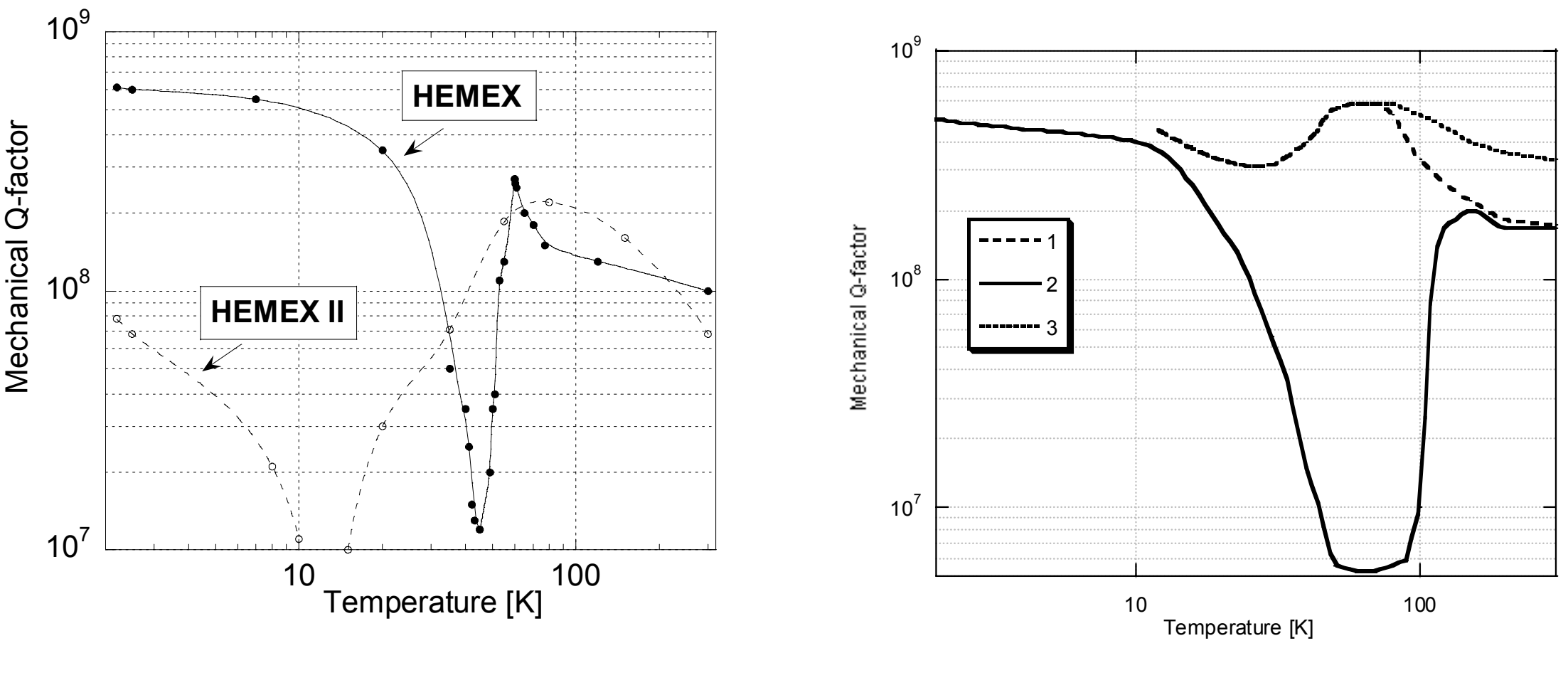

*Figure 6: Comparison of mechanical Q-factors of the HEMEX used in this work (first graph) against the sapphire used in by Braginsky[6] (second graph.) In the second graph the influence of spalling is shown; curve 1 is before spalling, curve 2 after spalling, and curve 3 is after high temperature annealing.*

The loss peak exhibited at 45K is thought to be due interacting combinations of point defects and dislocations brought about by cold working. Such a loss peak is known as the Bordoni peak. The Bordoni peak is (i) independent of strain amplitude, (ii) reproducible, (iii) can be observed in single crystals, and (iv) impurities are not necessary for its appearance. Measurements of this particular sample satisfy all four characteristics. High temperature (~2000°C) annealing of the sapphire is necessary if this loss peak is to be removed. We expect the intrinsic mechanical Q-factor to then increase to around $10^{10}$ at 4K after annealing.
Another white-purity sapphire cylinder from Crystal Systems was investigated. This piece (dubbed HEMEX II in this work to distinguish it from the MST) is identical in all regards save its diameter is 25mm rather than 50mm. At room temperature the measured mechanical Q-factor is $1.7 \times 10^7$, at 77K is $2.2 \times 10^8$, and at 4K is $7.5 \times 10^7$. In comparison to the larger piece, HEMEX II has a similar mechanical quality factor at room temperature and is slightly better around 77K. However, the Bordoni loss peak degrades the Q-factor at temperatures below 30K.

### *2.4. Requirements to measure the noise fluctuations of the SQL*

Assuming a simple filter, Figure 7 shows the RMS fluctuations of the thermal noise (narrow band) and the pump oscillator phase noise (broad band) with respect to the SQL in units $dB_{SQL}=20\log_{10}[\langle x\rangle/\langle x_{SQL}\rangle]$. From the intersections of narrow-band and broad-band lines we can identify the necessary requirements for the MST to measure the SQL. A mechanical Q-factor of $6 \times 10^8$ has been achieved in the MST, thus if we were to measure the SQL we would need a spectral purity of $10^{-21}$m/√Hz.
The phase noise of the microwave source is –137 dBc/Hz at 50 kHz, which equates to a spectral sensitivity limit due to phase noise of $4.2 \times 10^{-15}$m/√Hz. This level of noise renders measurements of the SQL impossible. However, it would be possible to implement a specifically designed liquid nitrogen cooled sapphire oscillator with a phase noise of –165 dBc/Hz, corresponding to a spectral sensitivity of $10^{-16}$m/√Hz. Further improvements can be realised using recent advances in oscillator technology[7],[8], which allow us to reach a spectral sensitivity of $10^{-19}$ m/√Hz.

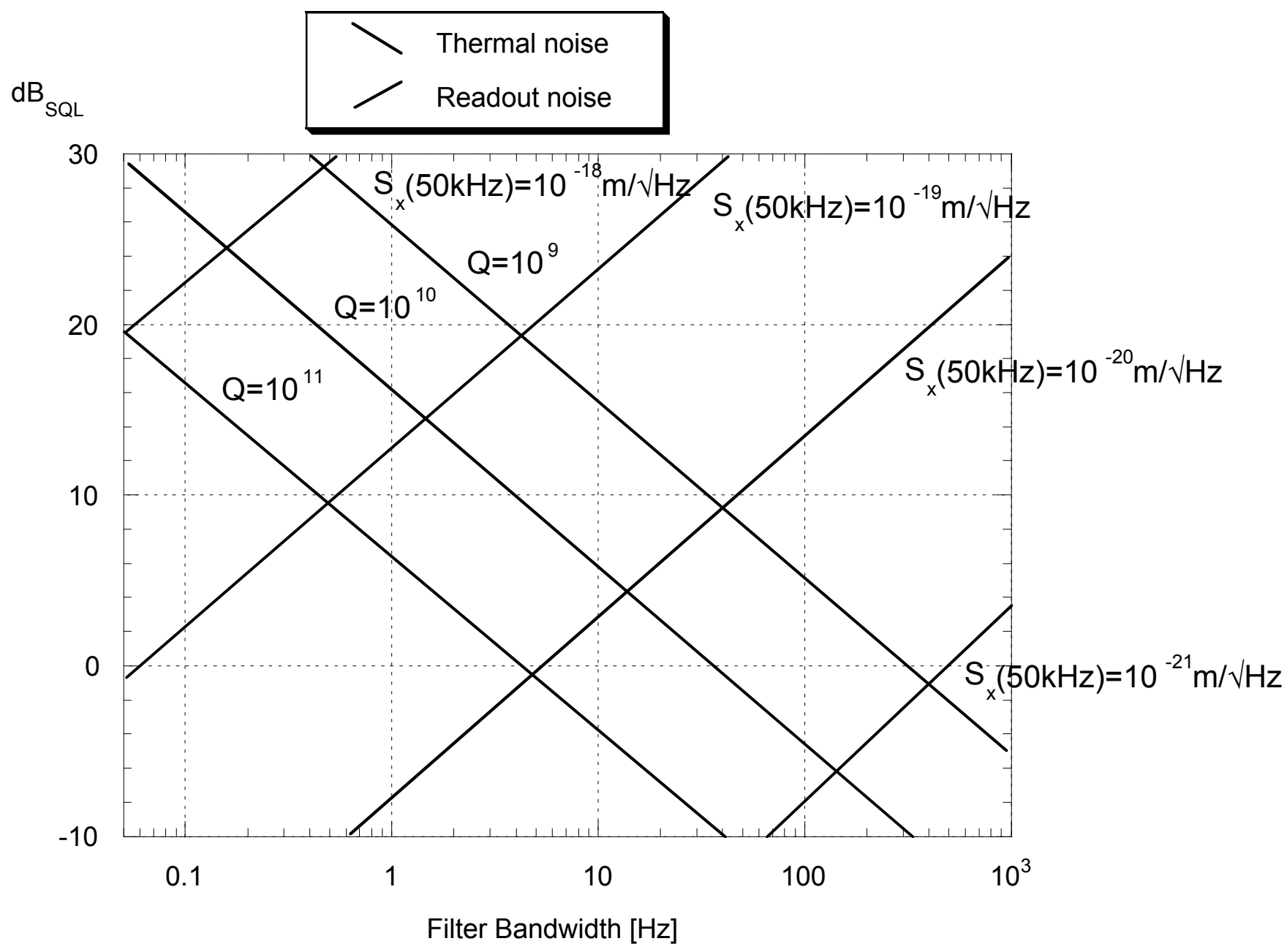

*Figure 7: Narrow-band and broad-band noise components of the MST as a function of sampling bandwidth ($1/t_{meas}$). The intersection of the lines gives the optimum bandwidth for a given thermal (Equation 19) and readout (Equation 21) noise.*

Given that it may be possible to achieve this spectral purity, the phase noise now needs to be suppressed by 40dB in order to measure the SQL. This is possible using an interferometric configuration, as shown in [9]. Such a configuration is shown schematically in Figure 8.

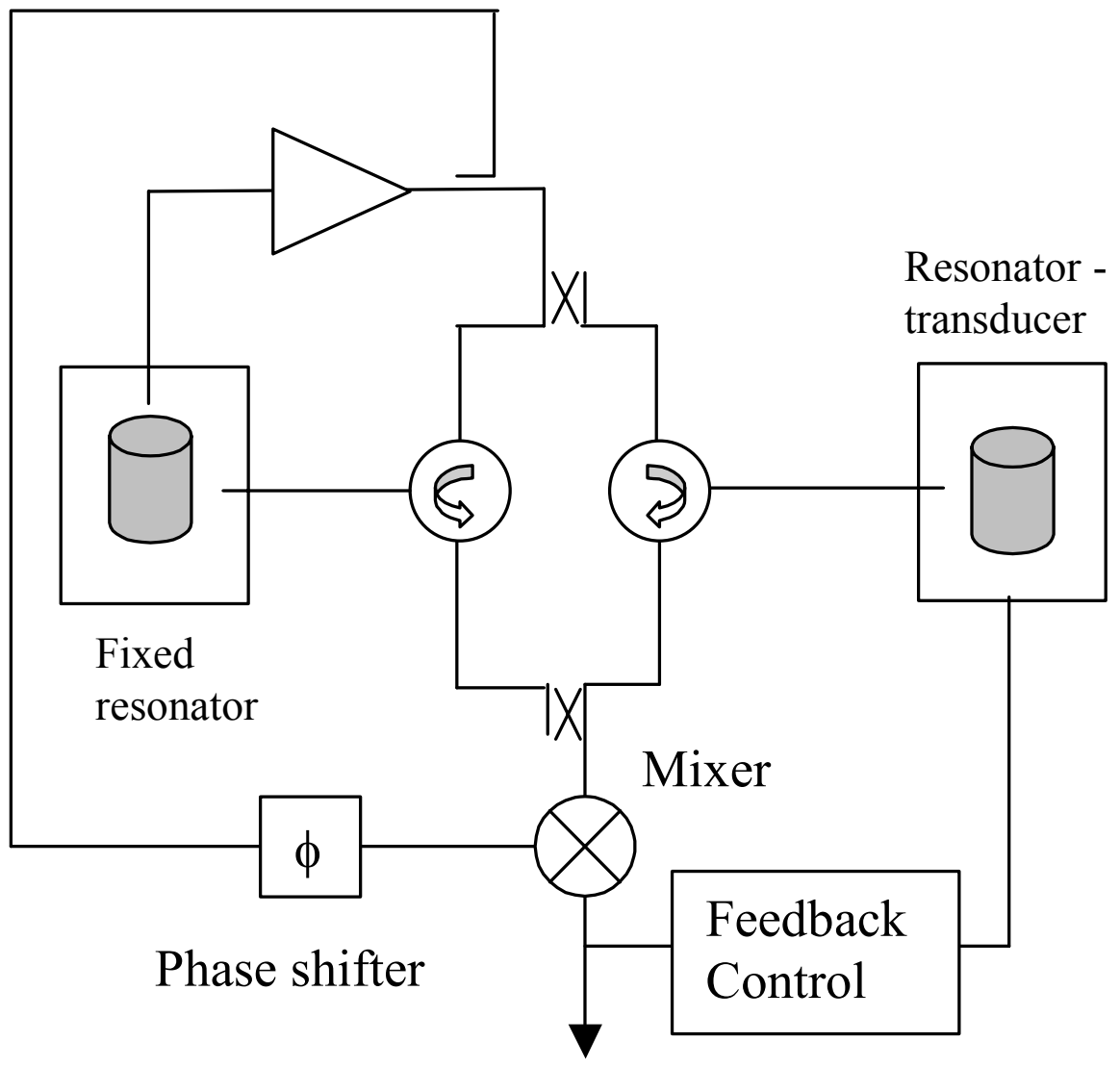

*Figure 8: An interferometric configuration; one resonator is a transducer, the other is used doubly in a pump oscillator and one arm of the interferometric configuration.*

Two sapphire dielectric resonators are used; one is a sapphire dielectric resonator transducer, and the other is a fixed resonator of similar dimensions. The fixed resonator is used twice;

firstly as a loop oscillator to provide a stable microwave frequency signal, and secondly in one arm of the interferometer.

A good balance requires the two resonators have approximately the same product $\beta Q$. If the interferometric system is fully balanced, all the pump phase noise is suppressed. In [9] a phase noise suppression of 26dB was achieved.

Another means of reaching the SQL would be to improve the mechanical Q-factor of the acoustic oscillator. As the intrinsic mechanical Q-factor of the MST has already been reached we would need another sapphire sample of higher purity and containing fewer dislocations. If this were used we could optimistically expect a mechanical Q-factor of $10^{11}$. This would relax the requirement on the spectral purity to $10^{-20}$ m/√Hz. In this case, only 20dB suppression of phase noise would be required in order to measure the SQL.

## 3. Conclusion

An effective suspension system has been developed and refined. A single loop of niobium wire of radius 125μm was found to provide optimal stability and isolation. The highest measured mechanical Q-factors recorded in this work are 1.4 x $10^8$ and 6 x $10^8$ at room temperature and 4K respectively.

The current sensitivity of the MST is of order $10^{-15}$ m/√Hz. This limit is set by the phase noise of the pump. The implementation of a specifically designed oscillator and using recent advances in oscillator technology should achieve a spectral sensitivity of $10^{-19}$m/√Hz. If a higher mechanical Q-factor was attained the MST has the potential to reach the goal of measuring the standard quantum limit in a macroscopic object.